\documentclass[aps,prl,reprint,showpacs]{revtex4-1}
\usepackage{graphicx}
\usepackage{color}
\usepackage{bm}
\usepackage{braket}
\usepackage{amsmath, amsthm, amssymb,mathrsfs}
\usepackage{csquotes}
\usepackage{soul}
\usepackage[colorlinks=true, citecolor=blue,allcolors=blue]{hyperref}

\DeclareMathOperator*{\SumInt}{%
\mathchoice%
  {\ooalign{$\displaystyle\sum$\cr\hidewidth$\displaystyle\int$\hidewidth\cr}}
  {\ooalign{\raisebox{.14\height}{\scalebox{.7}{$\textstyle\sum$}}\cr\hidewidth$\textstyle\int$\hidewidth\cr}}
  {\ooalign{\raisebox{.2\height}{\scalebox{.6}{$\scriptstyle\sum$}}\cr$\scriptstyle\int$\cr}}
  {\ooalign{\raisebox{.2\height}{\scalebox{.6}{$\scriptstyle\sum$}}\cr$\scriptstyle\int$\cr}}}
  \newcommand{\XUV}{{\scriptscriptstyle\mathrm{XUV}}}
\newcommand{\IR}{{\scriptscriptstyle\mathrm{IR}}}
\newcommand{\LCP}{{\scriptscriptstyle\mathrm{LCP}}}
\newcommand{\RCP}{{\scriptscriptstyle\mathrm{RCP}}}

\newcommand{\APT}{{\scriptscriptstyle\mathrm{APT}}}

\begin{document}

\title{Circular holographic ionization-phase meter}
\author{S.~Donsa$^{1*}$, N.~Douguet$^{2}$, J.~Burgd\"{o}rfer$^1$, I.~B\v rezinov\'a$^1$, and L.~Argenti$^{2,3*}$}
\affiliation{$^1 $Institute for Theoretical Physics, Vienna University of Technology, A-1040 Vienna, Austria, EU}
\affiliation{$^2$Department of Physics, University of Central Florida, Orlando, Florida 32186, USA}
\affiliation{$^3$CREOL, University of Central Florida, Orlando, Florida 32186, USA}
\date{\today}
\email{stefan.donsa@tuwien.ac.at}
\email{luca.argenti@ucf.edu}

\begin{abstract}
We propose an attosecond XUV-pump IR-probe photoionization protocol that employs pairs of counter-rotating consecutive harmonics and angularly resolved photoelectron detection, thereby providing direct measurement of ionization phases.  The present method, which we call circular holographic ionization-phase meter (CHIP), gives also access to the phase of photoemission amplitudes of even-parity continuum states from a single time-delay measurement, since the relative phase of one- and two-photon ionization pathways is imprinted in the photoemission anisotropy.
The method is illustrated with \emph{ab initio} simulations of photoionization via autoionizing resonances in helium.
The rapid phase excursion in the transition amplitude to both the dipole-allowed $(2s2p)^1$P$^{\rm o}$ and the dipole-forbidden $(2p^2)^1$D$^{\rm e}$ states are faithfully reproduced.
\end{abstract}

\maketitle
Excitation, ionization, and charge-migration events triggered by the absorption of light are at the core of several physical, chemical and biological processes of practical and conceptual importance, such as the photoelectric effect and the harvesting of light energy by natural and synthetic photoreceptors~\cite{Sch2017}. 
These processes rely on the excitation or ionization of electrons and take place on a sub-femtosecond timescale~\cite{LepIvaVrak2014}.
Time-resolved observation of their underlying dynamics became only accessible with the recent development of attosecond laser techniques~\cite{Krausz2009}
and progress in time and energy resolution~\cite{Sansone2006,Chang2007,Sansone2011,Popmintchev2012,kfir2015,Fleischer2014,Ferre15,Huang18,Fan15}. 
Attosecond pump-probe schemes such as streaking \cite{Kienberger2004,Itatani2002} and RABBITT (Reconstruction of Attosecond Beating By Interference of Two-photon Transition) \cite{Paul2001a} have made it possible to study ionization processes on their natural time-scale, and in particular to estimate the delay in the photoelectron emission from valence shells in atoms~\cite{Schultze2010a,Klunder2011a,Isinger17,OssRieNep2018} and molecules~\cite{Comby16,Beaulieu16,Beaulieu17,Serov16,Cirelli18}, as well as the time-resolved decay of metastable states~\cite{Argenti2010,WanChiChe2010,JimenezGalan2016,Gruson734,KalBlaSto2016,douguet18}.
These techniques, however, rely either on multi-photon transitions exclusively, or on ponderomotive energy shifts, and hence  offer only indirect access to the one-photon ionization phase, which is the key quantity to determine photoionization time delay.

\begin{figure}[t]
\includegraphics[width=\columnwidth]{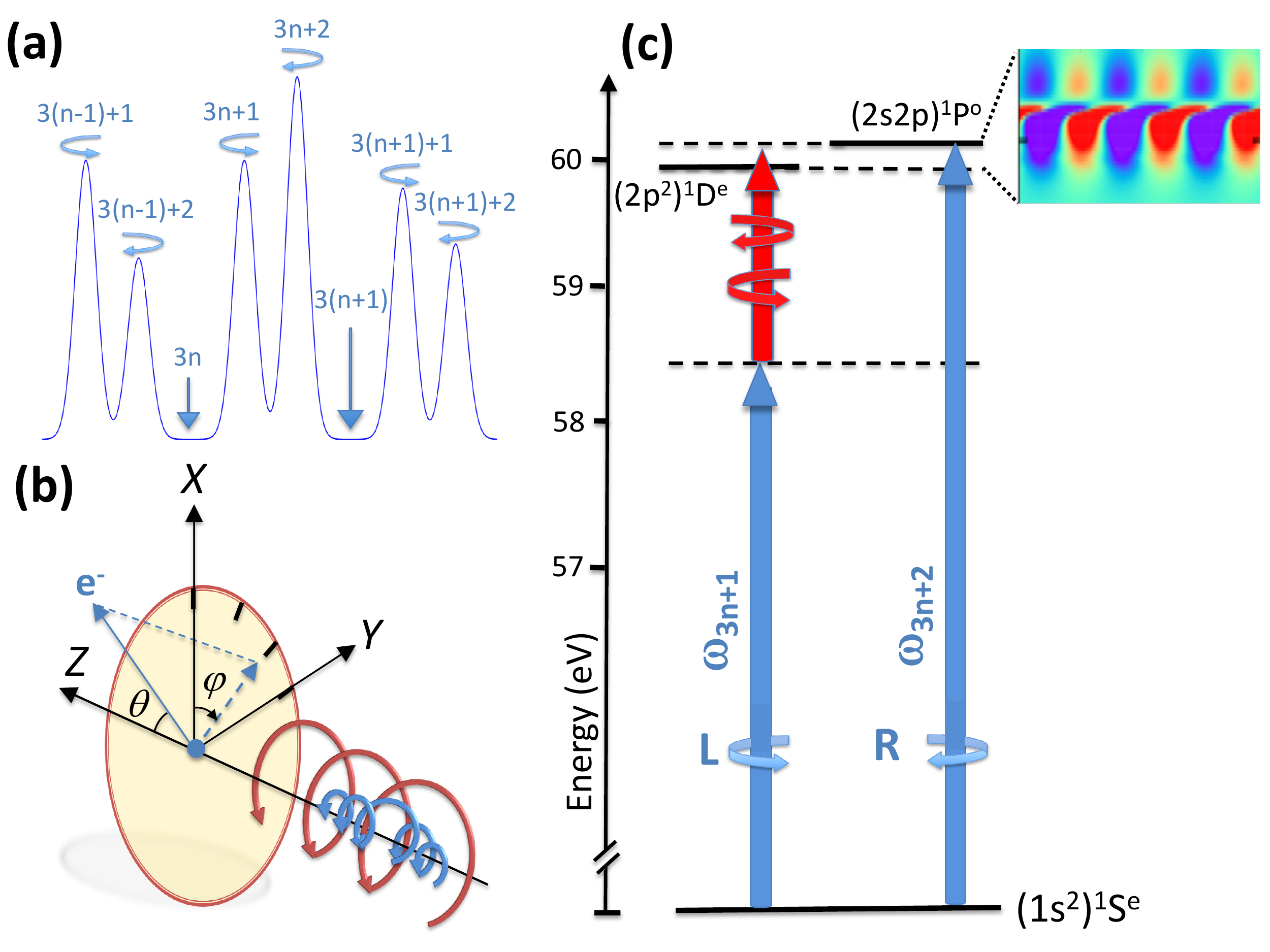}
\caption{(a) Typical spectrum of an APT generated by counter-rotating bichromatic fields, featuring isolated consecutive pairs of counter-rotating IR harmonics, $\omega=(3n+k)\omega_\IR$, $k=1,2$~\cite{kfir2015,Fan15}. 
(b) Geometry of the angularly resolved detection in CHIP.
(c) Principle of CHIP (the IR probe is either co- or counter-rotating with the lower harmonic in the pair), and an example of photoelectron momentum distribution (insets) as a function of the emission angle (horizontal axis) and energy (vertical axis), for the co-rotating case with the upper harmonics populating the $(2s2p)^1$P$^o$ helium resonant state. }
\label{fig:Fig1-bis2}
\end{figure}

In conventional RABBITT, a linearly polarized weak probe infrared (IR) pulse together with a pump attosecond-pulse train (APT) formed by odd-harmonics of the fundamental IR frequency $\omega_{\rm{IR}}$~\cite{Corkum1993} ionize the target. 
The ionization amplitudes resulting from the absorption of either harmonic $2n-1$  of the XUV pulse and one IR photon, or of harmonic $2n+1$ with the emission of one IR photon, interfere, giving rise to a sideband ionization signal that beats at twice the IR frequency as a function of the pump-probe delay $\tau$. 
In the absence of chirp~\cite{Veniard1996}, the beats phase offset $\Delta \delta$ corresponds to the phase difference between the two interfering ionization paths, and hence its variation with energy, $\partial \delta/\partial E\approx \Delta \delta / 2\omega_{\IR}$, approximates the group delay for the ionization by one XUV and one IR pulse. When the distortion by the IR is small, or can be separately accounted for, also the one-photon ionization delay can be reconstructed. Unambiguously extracting the one-photon phase from these measurements, however, is not always possible~\cite{JimenezGalan2016}. Recently, it has been demonstrated that bright, phase-matched, circularly-polarized high XUV harmonics can be produced by illuminating an active medium with intense bi-chromatic counter-rotating circularly-polarized laser pulses~\cite{kfir2015,Fan15}. The resulting high harmonic spectrum comprises counter-rotating isolated consecutive multiples $3n+1$ and $3n+2$ of the fundamental IR frequency, while the signal of the $3n$ harmonics is strongly suppressed (Fig.~\ref{fig:Fig1-bis2}~a). As shown in the following, these new bi-circular APT open the way to overcome the limitations of conventional interferometric methods.

In this letter we present the circular holographic ionization phase meter (CHIP), a pump-probe protocol, based on bi-circular APT, that allows us to directly retrieve the energy-resolved phase of either the one- or two-photon ionization amplitudes. The first key aspect of this method is the holographic read-out of a rapidly varying ionization phase in one arm of the interferometer, e.g., in proximity of a Fano resonance, relative to a smooth or nearly constant reference phase of the second arm.
The second key ingredient of CHIP is the use of a bi-circular APT in combination with a co- or counter-rotating IR pulse (Fig.~\ref{fig:Fig1-bis2}~b).
In CHIP the photoelectron phase is encoded in the photoemission angle instead of a time-delay scan of  the signal beating as in streaking and RABBITT.
The proposed scheme requires high-resolution measurement of the angular resolved photoelectron energy distribution (PED) as recorded by, e.g.  COLTRIMS detection~\cite{Ullrich2003a}. 

CHIP involves a single pair of consecutive circular harmonics $3n+k$ ($k=1,2$) of the XUV APT (Fig.~\ref{fig:Fig1-bis2}~a,b) with opposite helicity.
Assuming, e.g., the helicity of the lower harmonic $3n+1$ to be positive (or left-circularly polarized LCP) and the helicity of the upper harmonic $3n+2$ to be negative (or right-circular RCP), the electric field for light propagating in positive $\hat{z}$ direction is given by $E_k(t)=E_0[\cos(\omega_{k} t)\hat{x}+\nu_k\sin(\omega_{k} t)\hat{y}]$, $\nu_k=(-1)^{k+1}$ is the light helicity and $\omega_{k}=(3n+k)\,\omega_{\IR}$.
The IR probe pulse can be chosen to be either co-rotating ($\nu_{\IR}=+1$) or counter-rotating ($\nu_{\IR}=-1$) with the lower harmonic ($k=1$), allowing to control the interference between one- and two-photon amplitudes.

Employing \emph{ab initio} simulations for helium, a prototypical atomic system exhibiting Auger decay for which virtually exact numerical solutions of the time-dependent Schr\"odinger equation (TDSE) are possible, we demonstrate that the photoelectron angular distribution faithfully reproduces the energy dependence of the one-photon ionization amplitude of the atom near autoionizing resonances.
The emergence of different interference patterns and angular distributions can be explained within lowest-order perturbation theory (LOPT) [see supplementary material (SM)].
One-photon ionization of helium from its ground-state by the upper harmonic ($k=2$) creates a continuum wave at an energy $E=\left(3n+2\right)\omega_{\IR}-I_p$ with angular momentum quantum numbers $\ell=1,m=-1\sim Y_{\ell=1}^{m=-1}\left(\theta,\varphi\right)$.
The same energy can be reached via a two-photon absorption involving the lower harmonic $(k=1)$ and an IR photon.
If the IR is co-rotating with the lower APT harmonic, the angular dependence of the continuum wave populated by the two-photon path is just $Y_{2}^{2}\left(\theta,\varphi\right)$. 
The inference between the one-photon and two-photon paths is proportional to $\left|a^{(1)}_{1\,-1}  Y_{1}^{-1}\left(\theta,\varphi\right)+ a^{(2)}_{2\,2}  Y_{2}^{2}\left(\theta,\varphi\right) \right|$, where $a^{(1)}_{\ell m}$ and $a^{(2)}_{\ell m}$ are the one-photon and two-photon amplitudes, respectively. This interference, therefore, depends on the azimuthal angle $\varphi$ as $\cos \left(3 \varphi -\Delta_{\LCP}\right)$, where $\Delta_{\LCP}$ is the relative phase between the one- and the two-photon amplitude, 
\begin{equation}
\Delta_{\LCP}(E)=\delta^{(1)}(E) -\delta^{(2)}_{\LCP}(E).
\end{equation}
With a counter-rotating IR photon, the two-photon ionization populates a superposition of $s$ and $d$ waves with $m=0$, and hence the interference term, $\left| a^{(1)}_{1\,-1}  Y_{1}^{-1}\left(\theta,\varphi\right) + a^{(2)}_{0\,0} Y_{0}^{0}\left(\theta,\varphi\right) + a^{(2)}_{2\,0}  Y_{2}^{0}\left(\theta,\varphi\right)\right|$, depends on $\varphi$ as $\sim \cos \left( \varphi - \Delta _{\RCP}\right)$.
In CHIP, therefore, the interference between waves with $\ell=0,\,1,\,2$, gives access to the ionization amplitudes for dipole-forbidden or mixed-parity states.
Thanks to the APT short duration, a single photoelectron spectrum is sufficiently broad to record the full excursion of the scattering phases across a resonance. Details of the measurement for an IR with finite duration are discussed in the SM.
The interference between two different paths to two different partial waves is key to the holographic mapping of a single ionization phase.
If the phases of the two-photon transition are almost constant across a dipole-allowed resonance, for example, they can serve as holographic reference to measure a rapidly varying one-photon ionization phase $\arg[a^{(1)}_{1m}]$. 

We illustrate the capabilities of this concept with \emph{ab~initio} simulations of two prototypical examples (Fig.~\ref{fig:Fig1-bis2}~c).
The one-photon ionization phase near the dipole allowed doubly excited resonance   $(2s2p)^1$P$^{\rm o}$  and the ionization phase for the dipole-forbidden  $(2p^2)^1$D$^{\rm e}$ resonance of helium.
\begin{figure*}[!]
\includegraphics[width=16cm]{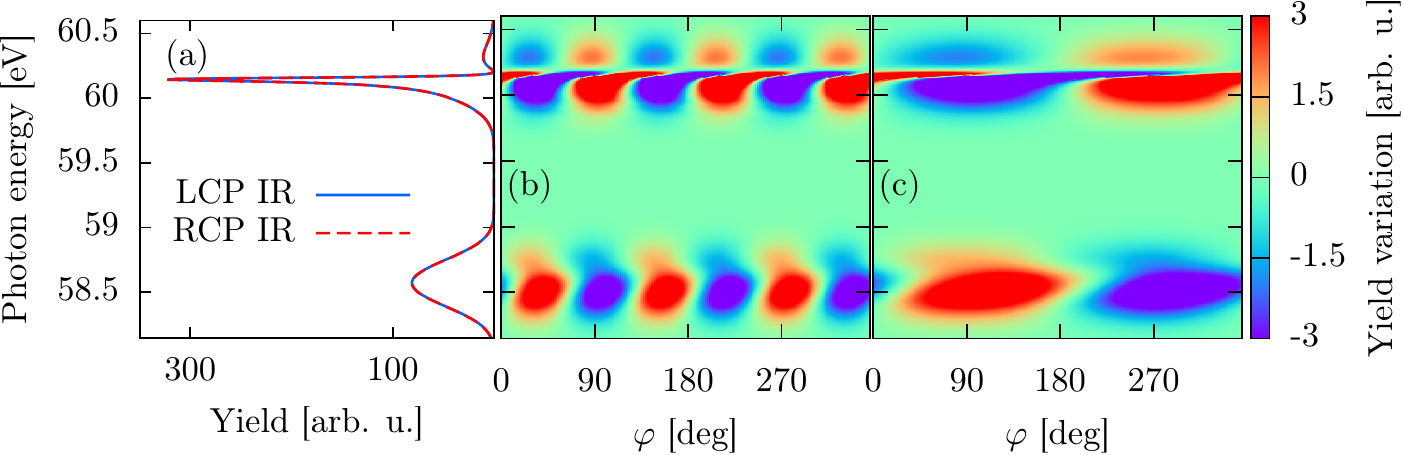}
\caption{Single ionization spectrum of atomic helium ionized by a bi-circular  APT and a co- or counter-rotating IR pulse ($I_{\APT}=10^{11} \mathrm{W}/\mathrm{cm^2}$ and $5 \cdot 10^{11} \mathrm{W}/\mathrm{cm^2}$ for the even and odd harmonics, respectively, $I_{\IR}=10^{9} \mathrm{W}/\mathrm{cm^2}$, $\tau_{\IR}\approx$ 13~fs, $\tau_{\APT}\approx$ 5~fs, $\theta=\pi/2$). 
(a) Spectrum at fixed azimuthal angle $\varphi=0^{\circ}$ in the polarization plane $\theta=\pi/2$.
Angular resolved photoemission spectrum with IR pulse (b) co-rotating with lower harmonic and (c) counter-rotating with the lower harmonic.
A constant background was subtracted to enhance the contrast.} \label{fig:osc_783nm}
\end{figure*}
We solve the full two-electron TDSE numerically from first principles \cite{Feist2008,DonBreNi2019} within the dipole approximation and in velocity gauge using the time-dependent close-coupling expansion (see SM for details).
To retrieve the phase modulation of the $(2s2p)^1$P$^{\rm o}$ resonance, we employ an APT that comprises $3n+1$ LCP and $3n+2$ RCP harmonics of the fundamental wavelength $\lambda_{\IR}$= 783 nm, with $n=11$, 12, and 13, as shown in Fig.~\ref{fig:Fig1-bis2}~(b), with intensity $10^{11} \mathrm{W}/\mathrm{cm^2}$ and $5 \cdot 10^{11} \mathrm{W}/\mathrm{cm^2}$ for the even and odd harmonics, respectively. 
We explore a wide range of IR intensities, from $I_{\IR}=10^{9} \mathrm{W}/\mathrm{cm^2}$, at which LOPT makes accurate predictions, to moderately strong fields $I_{\IR}=10^{11} \mathrm{W}/\mathrm{cm^2}$, typical in streaking settings, at which the interference contrast is strongly enhanced albeit deviations from LOPT estimates appear.
For measurements of the one-photon ionization phase the 38$^{{\rm th}}$ harmonic is tuned to  the optically allowed $(2s2p)^1$P$^\mathrm{o}$ resonance at 60.15~eV. 
As expected, whereas the photoelectron energy distributions of the non-resonant 37$^{{\rm th}}$ harmonic and of the  38$^{{\rm th}}$ harmonic resonant with the $(2s2p)^1$P$^\mathrm{o}$ resonance for the co- and counter-rotating IR pulses are very close in the perturbative limit, as shown for $\varphi=0^{\circ}$ in Fig.~\ref{fig:osc_783nm}a, their variations with $\varphi$ are drastically different, featuring a $\cos \left(3\varphi - \Delta_{\LCP}\right)$ and a $\cos \left(\varphi - \Delta_{\RCP}\right)$ dependence, respectively (see Fig.~\ref{fig:osc_783nm}b,c). At moderate IR intensities $I_\IR$ the angular-modulation fringes become clearly visible even without subtracting the average signal  (Fig.~\ref{fig:osc_783nm_int}) since their contrast increases linearly with the field strength (or $I_\IR^{1/2}$). The angular-beating phase $\Delta(E)=\delta^{(1)}\left(E\right)-\delta^{(2)}\left(E\right)$, which contains the ionization phases, can be readily retrieved by Fourier analysis of the signal  (Fig.~\ref{fig:osc_783nm}b,c).
\begin{figure}[b]
\includegraphics[width=\columnwidth]{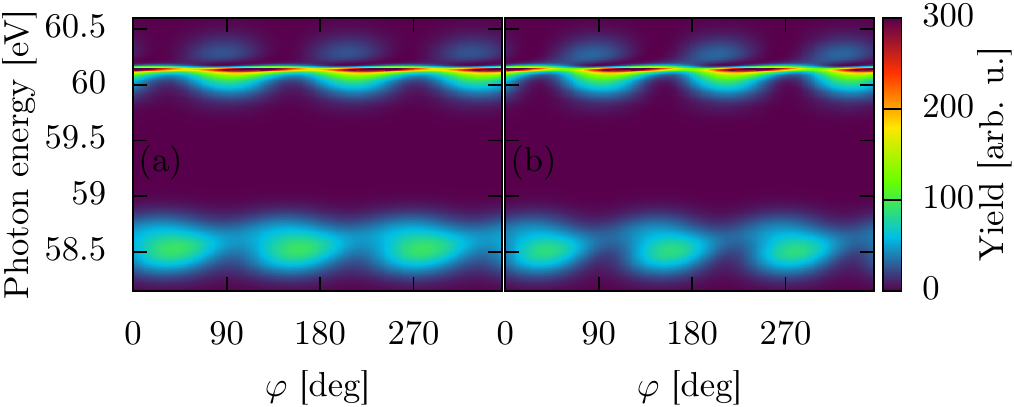}
\caption{Same as Fig.~\ref{fig:osc_783nm}b but without background subtraction and (a) $I_{\IR}=5 \times 10^{10}$ W/cm$^2$, (b) $I_{\IR}=1 \times 11^{11}$ W/cm$^2$.
} \label{fig:osc_783nm_int}
\end{figure}
The phase $\delta^{(1)}\left(E\right)$ across the $(2s2p)^1$P$^\mathrm{o}$ resonance at $E=60.15$~eV (Fig.~\ref{fig:phase_783nm}) exhibits the characteristic strong excursion with a discontinuous jump predicted by Fano's theory for resonant one-photon transitions, 
\begin{equation}\label{eq:Fano_phase}
\delta^{(1)}\left(E\right) = \delta^{(1)}_{\rm{bg}}(E)+\arg \left[(\epsilon+q)/(\epsilon + i)\right],
\end{equation}
where $\delta^{(1)}_{\rm{bg}}$ is a smooth background phase, $\epsilon=2\left(E-E_R\right)/ \Gamma$ is the relative energy distance from the resonance position $E_R$ in units of resonance's half width, $\Gamma/2$, and $q$ is the Fano asymmetry parameter~\cite{Fano1961}.
As long as the reference phase $\delta^{(2)}\left(E\right)$ is approximately constant across the resonance, which is indeed the case here (Fig.~\ref{fig:phase_783nm}), the angular-beating offset coincides with the ionization phase $\delta^{(1)}\left(E\right)$, up to an overall constant.
Indeed, our simulations convincingly demonstrate that the Fano phase of the optically allowed $(2s2p)^1$P$^\mathrm{o}$ resonance can be directly retrieved from CHIP with a high degree of accuracy.
\begin{figure}[b]
\includegraphics[width=\columnwidth]{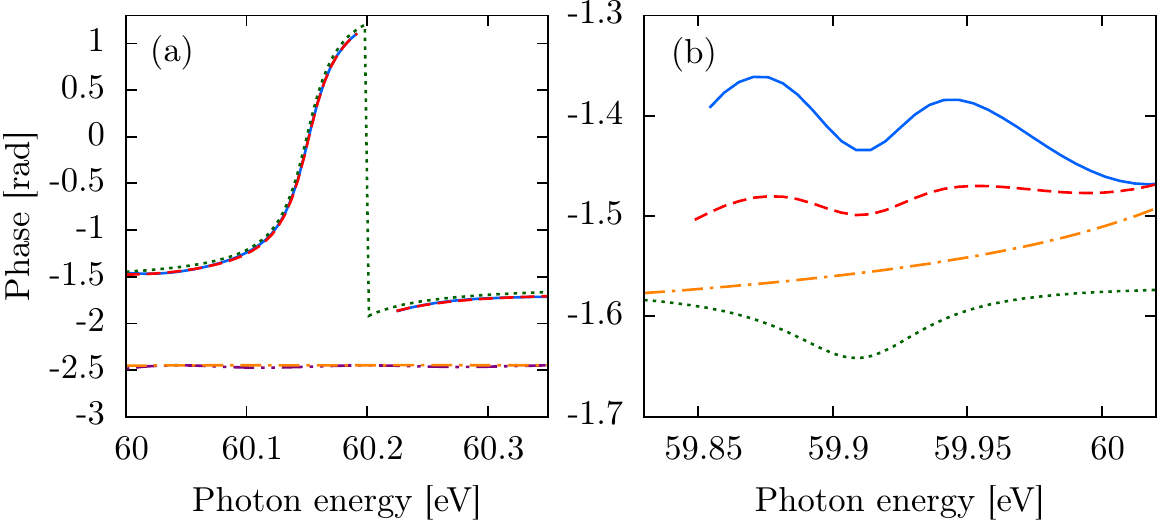}
\caption{\label{fig:phase_783nm} Ionization phases near the doubly-excited state resonances  $(2s2p)^1$P$^\mathrm{o}$ at $E=60.15$~eV and $(2p^2)^1$D$^\mathrm{e}$ at $E=59.91$~eV for IR pulse either co-rotating (blue solid) or counter-rotating (red dashed) with the lower harmonic $(3n+1)\omega_{IR}$.
(a) Comparison between simulation and the analytic prediction Eq.~\eqref{eq:Fano_phase} (green dotted) for the phase jump near the $(2s2p)^1$P$^\mathrm{o}$ resonance.
(b) Comparison between simulation and fit to the analytic prediction (green dotted) using $q\approx -0.07 + 0.99i$ for the two-photon excitation to  $(2p^2)^1$D$^\mathrm{e}$.
The smooth two-photon reference phase $\delta^{(2)}$ in (a) is shown for the $s$ (orange dashed-dotted) and $d$ (purple dashed-double dotted) partial wave separately.
The one-photon reference phase $\delta^{(1)}$ in (b) originates from the $p$ (orange dashed-dotted) partial wave.
See text for details.} 
\end{figure}
\\
We turn now to the dipole forbidden resonance $(2p^2)^1$D$^\mathrm{e}$ at $E=59.91$~eV which is simultaneously accessible within the bandwidth of the same harmonic of the ultrashort pulse (Fig.~\ref{fig:Fig1-bis2}c).
Here the two-photon transition involving the lower ($k=1$) APT harmonic and the IR photon scans the resonance while the one-photon transition by the higher ($k=2$) APT harmonic, far from the  $(2s2p)^1$P$^\mathrm{o}$ state, provides the nearly constant reference phase.
In contrast to the one-photon resonance, we observe a pronounced dependence of the two-photon resonance phase on the IR polarization. For the co-rotating case, the two-photon ionization path exclusively populates $^1$D$^{\rm e}$ states, whereas for the counter-rotating case both the resonant $\ell=2$ and the non-resonant $\ell=0$ states are populated, reducing the resonant contrast.
For the dipole-forbidden resonance, the ionization path does not exhibit the same excursion as the one-photon resonant scattering phase [Eq.~(\ref{eq:Fano_phase})].
This is because, in a two-photon process, the dipole transition to the final D resonant state $| 1s,E, \ell=2, m=2 \rangle$, proceeds from an intermediate P wave with outgoing character 
\begin{equation}\label{eq:two_photon_mat_element_d_res}
\begin{split}
&\langle 1s,E,2,2| T^{(2)} | g \rangle = \\
&  \langle 1s,E,2,2| D_1^{1} \SumInt \mathrm{d}E' \frac{  | 1s, E', 1,1 \rangle \langle 1s,E', 1,1 | D_1^{1} | g \rangle }{E-E'-\omega_{\IR}+ i \eta},
\end{split}
\end{equation}
which is not an eigenstate of the time-reversal operator and hence cannot be expressed as a purely real function \cite{JimenezGalan2016}. In \eqref{eq:two_photon_mat_element_d_res}, $D_1^{1}$ is the dipole operator for LCP.
Consequently, the Fano $q$ parameter for a resonance populated by a two-photon transition is inherently complex and the phase jump is blurred. 
Fitting the amplitude of the resonant ($\ell=$2,$m=$2) channel to $\left| (\epsilon+q)/(\epsilon+i)\right|^2$ (with $q \in \mathbb{C}$) yields $q \approx -0.07 + 0.99 i$, in line with the analytical prediction $q=i$~\cite{JimenezGalan2016} expected for long IR pulses.
The two-photon continuum state in the counter-rotating case is a superposition of an $s$ and a $d$ wave, and hence it is possible to disentangle their contribution as they have a known and different dependence on $\theta$. For example, since the $d$ wave vanishes at the magic angle $\theta_{\mathrm{m}}\approx54.74^{\circ}$, i.e., $Y_{2}^{0}(\theta_{\mathrm{m}},\varphi) =0$, CHIP  at $\theta_{\mathrm{m}}$ directly measures the $\ell=0$ ionization phase.
\begin{figure}[b]
\includegraphics[width=0.5\columnwidth]{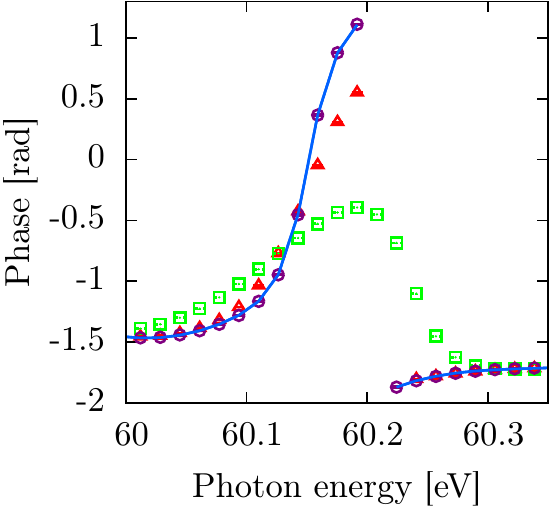}
\caption{Dependence of the retrieved ionization phase of the $(2s2p)^1$P$^\mathrm{o}$ Fano resonance on the angular and energy resolution $\Delta E$ of the detector. 
$-$ infinite resolution, $\circ$: reconstructed phase for only four values of $\varphi \in \left\{ 15^{\circ},60^{\circ},170^{\circ},240^{\circ}   \right\}$, $\triangle:$~$\Delta E=$~20~meV, $\square:$~$\Delta E=$~50~meV. See text for details.} \label{fig:EnergyConvolution}
\end{figure}

In CHIP, the angular modulation is inherently and rigorously sinusoidal. Measurements at only a few azimuthal angles (e.g., $\varphi=15^{\circ},60^{\circ},170^{\circ},240^{\circ}$), therefore, are sufficient to completely characterize it [Fig.~(\ref{fig:EnergyConvolution})].
Furthermore, the ionization phases can be reconstructed with an energy resolution that is limited only by that of the spectrometer but not by the spectral width of the IR.
To quantify the effect of the limited energy resolution on the extraction of ionization phases, we convolute the spectrum with instrumental resolutions of $20$~meV and $50$~meV width [Fig.~(\ref{fig:EnergyConvolution})]. 
At $20$~meV the phase excursion of the narrow $(2s2p)^1$P$^{\mathrm{o}}$ resonance is accurately reproduced, whereas at $50$~meV the phase jump is significantly smoothed.
The only main challenge in the implementation of CHIP, therefore, is to measure the angularly resolved photoelectron spectrum with a sufficiently high energy resolution. On the other hand, high-energy resolution is required only in the comparatively narrow width of a single harmonic, which can be easily achieved with a VMI detector in combination with tunable retardation plates~\cite{Rallis2014}.
Furthermore, while for the narrow $(2s2p)^1$P$^{\mathrm{o}}$ resonance a high spectral resolution is required, for broader resonances (e.g. in argon or xenon), the requirements on the spectral resolution are less demanding and hence CHIP is ideally suited to reconstruct their decay.

In the present \emph{ab initio} simulations we have used zero time delay ($\tau=0$) between ATP pump and IR probe. 
As long as the pump and probe overlap, however, any other time delay can be chosen, without altering the one-photon phase information. 
For monochromatic IR pulses a nonzero delay $\tau$ induces a trivial overall rotation of the angular distribution by $\Delta \phi=\omega_{\IR} \tau$.
Finite duration of the IR gives rise to an additional linear offset in the angular beating, on top of the global rotation. 
For Gaussian pulses, in particular, the phase depends linearly on the energy difference $\xi=E-\omega_k$ between the final electron energy and the harmonic central frequency,
$\partial^2\arg\mathcal{A}^{(2)}_{E}/\partial E\partial\tau\simeq -1/(1+\sigma_\XUV^2/\sigma_\IR^2)$,
where $\sigma_\XUV$ and $\sigma_\IR$ are the XUV and IR spectral widths (see SM). 
This effect, which is quantitatively confirmed in our simulations, can be easily accounted for in the analysis of experimental data where linear and even quadratic background (attochirp) are commonplace. Conversely, for finite pulses, the Fourier transform of the time-delay scan provides the complex two-photon matrix element.
Finally, in the co-rotating configuration with a $3\varphi$ angular dependence  the technique is insensitive to dipolar and quadrupolar distortions in the detection of the angular distribution due, e.g., to an imperfect beam alignment.  

To summarize, in this work we propose the circular holographic ionization phase meter (CHIP), a new attosecond spectroscopy based on bi-circular harmonics and angularly resolved photoelectron detection. 
The technique faithfully maps the phase of rapidly varying one- and two-photon transition amplitudes to the photoemission angle in the polarization plane, thereby avoiding the need of multiple time-delay measurements. 
We have illustrated the potential of this new method by directly ``measuring'' the Fano phase variation of the optically allowed $(2s2p)^1$P$^{\rm o}$ and the dipole forbidden $(2p^2)^1$D$^{\rm e}$ resonances in atomic helium. 
The method, thus, has the potential to give direct access to ionization phases of dipole forbidden and mixed-parity states in the continuum.

\section*{ACKNOWLEDGMENTS}
This work was supported by the United States National Science Foundation under NSF Grant No. PHY-1607588, by the WWTF through Project No. MA14-002, and  the  FWF  through  Projects  No.  FWF-SFB041-VICOM,  No. FWF-W1243-Solids4Fun. SD acknowledges support by the IMPRS-APS. Calculations were performed on the Vienna Scientific Cluster (VSC3).

\bibliography{biblio-new.bib}

\end{document}